\documentclass[amsmath,amssymb,prl,hyperlink,twocolumn,superscriptaddress,showpacs]{revtex4}

\usepackage{amsfonts}
\usepackage{dcolumn}
\usepackage{bm}
\usepackage{tikz}
\usepackage{color}
\usepackage[colorlinks=true,citecolor=blue,urlcolor=blue]{hyperref}
\usepackage{amsmath,amsfonts,amssymb,times,natbib}
\usepackage{multirow, makecell}
\usepackage[standard]{ntheorem}

\begin{document}

	\title{Experimental Test of Generalized Hardy's Paradox}

\author{Yi-Han Luo}
\affiliation{Hefei National Laboratory for Physical Sciences at Microscale and Department of Modern Physics, University of Science and Technology of China, Hefei, Anhui 230026, China}
\affiliation{CAS Centre for Excellence and Synergetic Innovation Centre in Quantum Information and Quantum Physics, University of Science and Technology of China, Hefei, Anhui 230026, China}

\author{Hong-Yi Su}
\affiliation{Graduate School of China Academy of Engineering Physics, Beijing 100193, People's Republic of China}

\author{He-Liang Huang}
\affiliation{Hefei National Laboratory for Physical Sciences at Microscale and Department of Modern Physics, University of Science and Technology of China, Hefei, Anhui 230026, China}
\affiliation{CAS Centre for Excellence and Synergetic Innovation Centre in Quantum Information and Quantum Physics, University of Science and Technology of China, Hefei, Anhui 230026, China}

\author{Xi-Lin Wang}
\affiliation{Hefei National Laboratory for Physical Sciences at Microscale and Department of Modern Physics, University of Science and Technology of China, Hefei, Anhui 230026, China}
\affiliation{CAS Centre for Excellence and Synergetic Innovation Centre in Quantum Information and Quantum Physics, University of Science and Technology of China, Hefei, Anhui 230026, China}

\author{Tao Yang}
\affiliation{Hefei National Laboratory for Physical Sciences at Microscale and Department of Modern Physics, University of Science and Technology of China, Hefei, Anhui 230026, China}
\affiliation{CAS Centre for Excellence and Synergetic Innovation Centre in Quantum Information and Quantum Physics, University of Science and Technology of China, Hefei, Anhui 230026, China}

\author{Li Li}
\affiliation{Hefei National Laboratory for Physical Sciences at Microscale and Department of Modern Physics, University of Science and Technology of China, Hefei, Anhui 230026, China}
\affiliation{CAS Centre for Excellence and Synergetic Innovation Centre in Quantum Information and Quantum Physics, University of Science and Technology of China, Hefei, Anhui 230026, China}

\author{Nai-Le Liu}
\email{nlliu@ustc.edu.cn}
\affiliation{Hefei National Laboratory for Physical Sciences at Microscale and Department of Modern Physics, University of Science and Technology of China, Hefei, Anhui 230026, China}
\affiliation{CAS Centre for Excellence and Synergetic Innovation Centre in Quantum Information and Quantum Physics, University of Science and Technology of China, Hefei, Anhui 230026, China}

\author{Jing-Ling Chen}
\email{chenjl@nankai.edu.cn}
\affiliation{Theoretical Physics Division, Chern Institute of Mathematics, Nankai University, Tianjin 300071, People's Republic of China}

\author{Chao-Yang Lu}
\affiliation{Hefei National Laboratory for Physical Sciences at Microscale and Department of Modern Physics, University of Science and Technology of China, Hefei, Anhui 230026, China}
\affiliation{CAS Centre for Excellence and Synergetic Innovation Centre in Quantum Information and Quantum Physics, University of Science and Technology of China, Hefei, Anhui 230026, China}

\author{Jian-Wei Pan}
\affiliation{Hefei National Laboratory for Physical Sciences at Microscale and Department of Modern Physics, University of Science and Technology of China, Hefei, Anhui 230026, China}
\affiliation{CAS Centre for Excellence and Synergetic Innovation Centre in Quantum Information and Quantum Physics, University of Science and Technology of China, Hefei, Anhui 230026, China}

\begin{abstract}
  Since the pillars of quantum theory were established, it was already noted that quantum physics may allow certain correlations defying any local realistic picture of nature, as first recognized by Einstein, Podolsky and Rosen. These quantum correlations, now termed quantum nonlocality and tested by violation of Bell's inequality that consists of statistical correlations fulfilling local realism, have found loophole-free experimental confirmation. A more striking way to demonstrate the conflict exists, and can be extended to the multipartite scenario. Here we report experimental confirmation of such a striking way, the multipartite generalized Hardy's paradoxes, in which no inequality is used and the conflict is stronger than that within just two parties. The paradoxes we are considering here belong to a general framework [S.-H. Jiang \emph{et al.}, Phys. Rev. Lett. 120, 050403 (2018)], including previously known multipartite extensions of Hardy's original paradox as special cases. The conflict shown here is stronger than in previous multipartite Hardy's paradox. Thus, the demonstration of Hardy-typed quantum nonlocality becomes sharper than ever.
\end{abstract}

    \pacs{03.65.Ud, 03.67.Mn, 42.50.Xa}

	\maketitle


\emph{Introduction.---}Quantum mechanics, according to Bell's theorem~\cite{bell64}, violates the notion of local realism that is always assumed true in the so called ``classical theories''. The conflict was first raised by Einstein, Podolsky and Rosen (EPR) back to 1935~\cite{epr}, then put down to physical ground by Bell in 1964, who showed it through quantum violation of an inequality bounded by the local hidden variable (LHV) theories~\cite{bell64}. Bell's inequality, however, uses statistics of joint measurements, while Hardy's paradox~\cite{hardy92,hardy93}, a logical quantum-versus-classical contradiction without the need of inequalities, provides a more striking conflict than that with inequalities. In particular, Hardy's paradox enables us to prove Bell's theorem under the same condition, proposed by EPR that one party's measurement outcome allows this party to predict the other party's measurement outcome, \emph{with certainty}.
Thus, Hardy's paradox, in any of its high-dimensional~\cite{chen2003} or multipartite extensions~\cite{cereceda2004}, giving a simple way to show that quantum correlation cannot be explained with classical theories, is arguably considered as ``the simplest version of Bell's theorem'' and ``one of the strangest and most beautiful gems yet to be found in the extraordinary soil of quantum mechanics''~\cite{mermin94}.

Hardy's original proof of nonlocality involved two particles, a positron and an electron in two overlapping Mach-Zehnder interferometers.
The paths of particles
are arranged such that the average probability of positron-electron annihilation does not equal one~\cite{Hardy1992}.
In fact, there are some remarkable merits in generalizing Hardy's paradox to the multipartite scenarios. First, differing from the GHZ paradox~\cite{GHZ89}, Hardy's paradox considers joint correlations in the form of probabilities, in a way that when it transforms to a group of correlation functions, the Hardy-typed Bell's inequality~\cite{Gill2008}, which is derived from the paradox, has not only fully correlated functions but also partially correlated ones --- i.e., the latter involve only a subset of parties, instead of involving all of them as in the former --- a vital property that makes the inequality capable of detecting more quantum states~\cite{chen08} (see also the results in~\cite{SG2001} for the Mermin-Ardehali-Belinskii-Klyshko inequality~\cite{Mermin1990,Ardehali,BK}). Second, with the GHZ state prepared, the quantum-classical conflict is sharper in the three- and four-party scenarios than in the two-party scenario~\cite{cereceda2004}. This feature increases the noise resistance in practical experiments, despite the difficulties of preparing multipartite entangled states. Third, only when $n\geq3$ is it possible to create the Hardy-typed paradox for mixed states~\cite{kar97}, extending the study of quantum paradoxes well beyond pure states in theory itself.

So far, a number of experiments have been carried out to confirm the paradox in two-particle systems~\cite{Hardy-Exp-1,Hardy-Exp-2,Hardy-Exp-3,Hardy-Exp-4,Hardy-Exp-5,Hardy-Exp-6,Hardy-Exp-7,Hardy-Exp-8,Hardy-Exp-9}. However, Hardy's paradox for multipartite systems have not been experimentally demonstrated, even for the three-qubit system. In this Letter, we perform the first experimental test of quantum nonlocality via a generalized Hardy's paradox proposed in~\cite{chen2018}. This paradox belongs to a general framework for $n$-particle Hardy's paradoxes, including Hardy's original one for two-qubit and Cereceda's extension for arbitrary $n$-qubit as special cases. Theoretically, it is proved that for any $n\geq 3$, the generalized paradox is always stronger than previously known ones. Furthermore, its corresponding Hardy-typed inequalities have a lower visibility with the GHZ state corrupted by white noise. The inequalities are \emph{tight} in that they coincide with the facets of LHV polytopes~\cite{tight}.

\begin{figure*}[!htp]
\setlength{\abovecaptionskip}{-2 mm}
\setlength{\belowcaptionskip}{-5 mm}
    \centering
        \includegraphics[width=\linewidth]{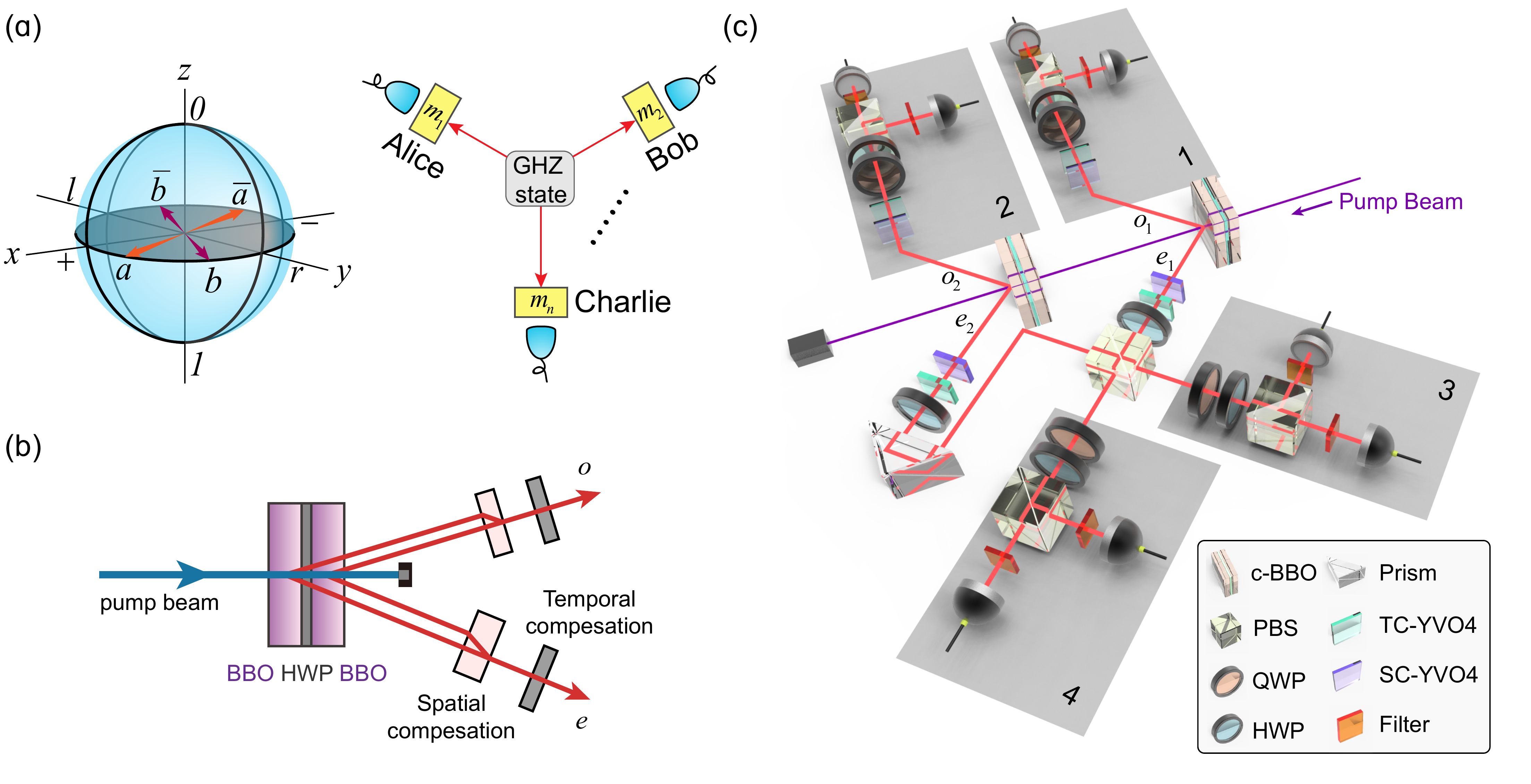}\vspace{3mm}
\caption{(a) Diagram of multipartite experiment to test local hidden-varible theories. Particles of GHZ state are distributed to $n$ spatially separated parties, local measurements are performed along some directions on Bloch sphere, respectively. In gerneralized Hardy's paradox, all projections are performed along vectors in the shaded plane, not limited to $x$ and $y$ axies,  which is different from GHZ paradox. (b) Sandwich-like BBO + HWP + BBO geometry for generating entangled photons. Birefringent crystals are used to compensate spatial and temporal displacements between two pairs of photons generated by the two single BBO. (c) Experimental setup.
The pump beam, with a central wavelength 394 nm, a duration of 140 fs and a repetition rate of 80 MHz, passes through two compound-BBOs (c-BBO) successively.
Photons in spatial modes $e_1$ and $e_1$ overlap on PBS, a movable prism is used to guarantee the temporal delay between two paths are equal. Four measurement modules are marked as 1 to 4 in the figure, each contains a QWP, a HWP and a PBS. All photons are spectrally filtered with 3-nm bandwidth filters. SC-YVO4: YVO4 crystal for spatial compensation; TC-YVO4: YVO4 crystal for temporal compensation.}\vspace{5mm}
\label{figsetting}
\end{figure*}


Consider an $n$-qubit system, in which qubits are labeled with the index set $I_n=\{1,2,\cdots,n\}$. For each qubit, e.g., the $k$-th, one can choose a projector from set $\{a_k,\bar{a}_k,b_k,\bar{b}_k\}$ to perform the measurement, where $a$ and $b$ denote two different projectors with binary outcome values $\{0,1\}$, the subscript indicates on which qubit the projection is performed, the barred projectors is orthogonal with the unbarred by defining $\bar a=\openone-a$ and $\bar b=\openone-b$, as illustration in left panel of figure \ref{figsetting} (a). For simplicity, we employ subscripts representing subsets of $I_n$, e.g., $\bar{a}_\alpha=\prod_{k\in\alpha}\bar{a}_k$, where $\alpha\subseteq I_n$ is a subset of all the $n$ qubits; we also employ $\bar{\alpha}=I_n/\alpha$. Moreover, we denote $|\alpha|$ as the size of the subset $\alpha$ --- i.e., the number of elements in $\alpha$ --- and successful probablity $p(a_1=1,a_2=1,\cdots)$ as $p(a_1a_2\cdots)$.

For any given $\alpha$ and $\beta$ satisfying $2\leq |\alpha|\leq n$, $1\leq|\beta|\leq|\alpha|$ and $|\alpha|+|\beta|\leq n+1$, in the regime of the LHV theory, the conditions
$$
p(b_\alpha a_{\bar{\alpha}})=p(\bar{b}_\beta a_{\bar{\beta}})=0,~~~\forall \alpha, \beta\subseteq I_n,
$$
must lead to
$$
p(a_{I_n})=0.
$$
In quantum theory, however, by choosing appropriate $a$ and $b$ for the GHZ state, $p(a_1a_2\cdots)$ can be greater than zero. Hence the quantum-classical conflict appears. This is the generalized Hardy's paradox~\cite{chen2018}.

For given $n, |\alpha|$ and $|\beta|$, let us label the generalized paradox as the $[n;|\alpha|,|\beta|]$-scenario. For instance, the previously known multipartite Hardy's paradox corresponds to the $[n;n,1]$-scenario~\cite{Cereceda}, with $p_{\rm std}(a_{I_3})=1/8$ and $p_{\rm std}(a_{I_4})=3/32$ for the 3- and the 4-qubit systems. Of all $[n;|\alpha|,|\beta|]$-scenarios, the optima of $p(a_{I_n})$ can be achieved by choosing $|\alpha|=|\beta|=2$, equal to $p_{\rm opt}(a_{I_3})=1/4$ and $p_{\rm opt}(a_{I_4})=1/8$, respectively. The Hardy-typed paradox is thus revealed more sharply than ever.

Diagram of generalized Hardy's paradox experiment is shown in right panel of figure \ref{figsetting} (a). Similar to other nonlocality tests, the GHZ state is distributed to $n$ spatially separated parties, with each particle being then measured along some direction represented on the Bloch sphere. Compared to the GHZ paradox,
there are mainly two different aspects to test generalized Hardy's paradox: (a)
the generalized Hardy's paradox involves the successful probabilities of projectors, other than expectations of Pauli's operators;
(b) all the projectors are in the $x$-$y$ plane of Bloch sphere (shaded plane in left panel), other than just along $x$ and $y$ axies. 



\emph{Experimental Setup and Results.---}Figure \ref{figsetting} (c) illustrates the setup of our experiment. To generate 3- and 4-photon GHZ state, we first prepare two pairs of polarized-entangled photons in the state $|\phi^+\rangle=(|H\rangle|H\rangle+|V\rangle|V\rangle)/\sqrt{2}$, then use a polarizing beam splitter (PBS) to combine them into 4-photon GHZ state.

To obtain polarized-entangled photons, we adopt beamlike type-II compound $\beta$-barium borate (c-BBO) crystal used in~\cite{10-photon}, where a half-wave plate (HWP) is sandwiched between two 2-mm-thick, identically cut BBOs, as shown in figure \ref{figsetting} (b). Pumped by ultraviolet (UV) laser pulses, photon pairs, emitted as two seperate circular beams of Gaussian intensity distribution, are generated through spontaneous parametric down conversion (SPDC)~\cite{rmp-photon} from the left BBO crystal. The polarization state can be written as $|H_o\rangle|V_e\rangle$, where $H~(V)$ denotes the horizontal (vertical) polarization state of photons and subscripts $o~(e)$ indicate two spatial modes on which photons are ordinary (extraordinary) light. Passing the HWP, above two-photon state is converted to $|V_o\rangle|H_e\rangle$, which is then superposed with photon pair $|H_o\rangle|V_e\rangle$ generated by the right BBO. By careful spatial and temporal compensations, polarized-entangled photons in the state $|\psi^+\rangle=(|H_o\rangle|V_e\rangle+|V_o\rangle|H_e\rangle)/\sqrt{2}~$  are prepared. We can convert $|\psi^+\rangle$ to desired $|\phi^+\rangle$ by inserting HWP in spatial mode $e$. In our experimental setup, pump pulses pass through two c-BBOs successively. In this way two pairs of photons are prepared in state $(|H\rangle_{o_1}|H\rangle_{e_1}+|V\rangle_{o_1}|V\rangle_{e_1})/\sqrt{2}~\otimes(|H\rangle_{o_2}|H\rangle_{e_2}+|V\rangle_{o_2}|V\rangle_{e_2})/\sqrt{2}$ with non-vanishing probability, here we use subscript $1$ and $2$ to distinguish the two c-BBOs from which photons generate.

The two photons in modes $e_1$ and $e_2$ are then combined on a PBS. The PBS transmits $H$ and reflects $V$, leading to a coincidence registration of a single photon at each output. In this way the two terms $|H\rangle^{\otimes 4}$ and $|V\rangle^{\otimes 4}$ are post-selected, the two entangled-photon pairs are therefore effectively projected into the four-photon GHZ state:
\begin{equation}
\begin{array}{r@{}l}
|G_4\rangle=\displaystyle\frac{1}{\sqrt{2}}(|H\rangle_{o_1}&|H\rangle_{e_1}|H\rangle_{o_2}|H\rangle_{e_2}\\
&+|V\rangle_{o_1}|V\rangle_{e_1}|V\rangle_{o_2}|V\rangle_{e_2}).
\end{array}
\label{4photon}
\end{equation}
To achieve good spatial and temporal overlap, a movable prism is used to guarantee the delay between two spatial modes are equal. Based on  state (\ref{4photon}), it is straightforward to generate 3-photon GHZ state projecting one of the photon (in this experiment the photon in spatial mode $o_1$) into $|+\rangle = (|H\rangle+|V\rangle)/\sqrt{2}$.

Subsequently, each photon of the GHZ states is led to a measurement module, which contains a quarter-wave plate (QWP), a half-wave plate (HWP) and a PBS. By setting QWP and HWP at $(0^\circ, 0^\circ)$ and $(45^\circ,22.5^\circ-\theta/4)$, measurement under basis $|H\rangle,|V\rangle$ and $(|H\rangle\pm e^{i\theta}|V\rangle)/\sqrt{2}$ can be performed. After the PBS, transmitted and reflected modes correspond to the two orthogonal projection of measurement basis. All the $16$ four-photon coincidence events can be simultaneously recorded by a coincidence counting system. To guarantee indistinguishability, each photon is spectrally filtered before entering detectors.

In this setup, the four-photon coincidence count rate is about $24.1$ Hz. To verify entanglement of the 3- and 4- photon GHZ state, we measured the entanglement witness~\cite{witness1,witness2}
\begin{equation}
W_{|G_N\rangle} = \frac{\openone}{2}-|G_N\rangle\langle G_N|,
\end{equation}
\begin{equation}
|G_N\rangle\langle G_N| = \frac{1}{2}P^N+\frac{1}{2N}\sum_{k=0}^{N-1}(-1)^kM_k,
\end{equation}
where population $P^N=|H\rangle\langle H|^{\otimes N}+|V\rangle\langle V|^{\otimes N}$, $M_k=[\cos(k\pi/N) \sigma_x +\sin(k\pi/N)\sigma_y]^{\otimes N}$. Expectations of $P^N$ and $M_k$ can be obtained by measurement under basis $|H\rangle,|V\rangle$ and $(|H\rangle\pm e^{ik\pi/N}|V\rangle)/\sqrt{2}$. Experimentally, we yield witness of 3- and 4-photon GHZ states as $-0.417(9)$ and
$-0.398(8)$, both are negative by more than 40 standard deviations, which prove presence of genuine 3- and 4-partite entanglement. From the witnesses, we can directly calculate fidelities of GHZ states as $0.917(9)$ and $0.898(8)$, which is defined as $F(\rho_{\exp})=\langle G_N|\rho_{\exp}|G_N\rangle$~\cite{Nielsen}.

\begin{table}[t]
\caption{\label{tab:table1}
Angles of HWP. HWP$_i~~(i=a,b)$ represents angle of HWP for projection $|i\rangle\langle i|$ and $|\bar{i}\rangle\langle \bar{i}|$.
}
\begin{ruledtabular}
\begin{tabular}{@{\hspace{5mm}}ccc@{\hspace{5mm}}cc@{\hspace{5mm}}}
\textrm{$n$}&
\textrm{$\alpha$}&
\textrm{$\beta$}&
\textrm{HWP$_a$}&
\textrm{HWP$_b$}\\
\colrule\\[-6pt]
3 & 3 & 1 & 30$^\circ$ & 7.5$^\circ$\\
3 & 2 & 2 & 22.5$^\circ$ & 0$^\circ$\\
4 & 4 & 1 & 26.2$^\circ$ & 11.25$^\circ$\\
4 & 2 & 2 & 22.5$^\circ$ & 0$^\circ$\\
\end{tabular}
\end{ruledtabular}
\end{table}
\vspace{3mm}

\begin{table*}
\caption{\label{tab:table2}
Experimental results. For the four scenari, from (a) to (d), all the projectors are listed in the table, together with experimentally measured expectations (successful probabilities) and standard deviations, standard deviations are calculated from Poissonian counting statistics of the raw detection events. Four theoritically non-vanishing probabilities are listed as bold font; all the measured probabilties with theoritical prediction zero are also listed in the table.
}
\begin{ruledtabular}
\begin{tabular}{cccccccccccc@{\hspace{8mm}}cc}
\multirow{3}*{(a)} & \multicolumn{2}{c}{$n=3$} & \multirow{3}*{(b)} & \multicolumn{2}{c}{$n=3$} & \multirow{3}*{(c)} & \multicolumn{2}{c}{$n=4$} & \multirow{3}*{(d)} & \multicolumn{4}{c}{$n=4$}\\
 & \multicolumn{2}{c}{$|\alpha|=3,|\beta|=1$} &  & \multicolumn{2}{c}{$|\alpha|=2,|\beta|=2$} &  & \multicolumn{2}{c}{$|\alpha|=4,|\beta|=1$} &  & \multicolumn{4}{c}{$|\alpha|=2,|\beta|=2$}\\[2pt]
\Xcline{2-3}{0.4pt}\Xcline{5-6}{0.4pt}\Xcline{8-9}{0.4pt}\Xcline{11-14}{0.4pt}\\[-8pt]
 & Projector & Probability & & Projector & Probability & & Projector & Probability & & Projector & Probability & Projector & Probability \\
\colrule\\[-6pt]
 & $b_1b_2b_3$ & 0.019(7) & & $a_1b_2b_3$ & 0.016(6) & & $b_1b_2b_3b_4$ & 0.005(3) & & $a_1a_2b_3b_4$ & 0.008(3) & $a_1a_2\bar{b}_3\bar{b}_4$ & 0.011(4) \\
 & $\bar{b}_1a_2a_3$ & 0.008(5) & & $a_1\bar{b}_2\bar{b}_3$ & 0.024(5) & & $\bar{b}_1a_2a_3a_4$ & 0.002(2) & & $a_1b_2a_3b_4$ & 0.004(2) & $a_1\bar{b}_2a_3\bar{b}_4$ & 0.010(3) \\
 & $a_1\bar{b}_2a_3$ & 0.032(9) & & $b_1a_2b_3$ & 0.007(4) & & $a_1\bar{b}_2a_3a_4$ & 0 & & $a_1b_2b_3a_4$ & 0.006(3) & $a_1\bar{b}_2\bar{b}_3a_4$ & 0.012(4) \\
 & $a_1a_2\bar{b}_3$ & 0.017(7) & & $\bar{b}_1a_2\bar{b}_3$ & 0.020(7) & & $a_1a_2\bar{b}_3a_4$ & 0.007(4) & & $b_1a_2b_3a_4$ & 0.007(3) & $\bar{b}_1a_2\bar{b}_3a_4$ & 0.009(3) \\
 &  &  & & $b_1b_2a_3$ & 0.014(6) & & $a_1a_2a_3\bar{b}_4$ & 0.005(3) & & $b_1a_2a_3b_4$ & 0.005(2) & $\bar{b}_1a_2a_3\bar{b}_4$ & 0.010(3) \\
 &  &  & & $\bar{b}_1\bar{b}_2a_3$ & 0.028(9) & &  &  & & $b_1b_2a_3a_4$ & 0.007(3) & $\bar{b}_1\bar{b}_2a_3a_4$ & 0.006(3) \\[5pt]
 & $\bm{a_1a_2a_3}$ & \textbf{0.132(16)} & & $\bm{a_1a_2a_3}$ & \textbf{0.259(22)} & & $\bm{a_1a_2a_3a_4}$ & \textbf{0.093(12)} & & $\bm{a_1a_2a_3a_4}$ & \textbf{0.127(12)} &  &  \\
\end{tabular}
\end{ruledtabular}
\end{table*}

To demonstrate generalized Hardy's paradox, two typical cases from generalized $[n;|\alpha|,|\beta|]$-scenario are selected. For 3- and 4-photon GHZ states, we both test the optimal case $|\alpha|=2,|\beta|=2$ and standard case $|\alpha|=n,|\beta|=1$ to reveal there exist optimal conditions in the regime of generalized Hardy's paradox, in which the probabilities $p(a_{I_n})$ surpass standard multipartite Hardy's paradox~\cite{Cereceda}. To make probabilities $p(b_\alpha a_{\bar{\alpha}})$ and $p(\bar{b}_\beta a_{\bar{\beta}})$ be zero, direction of the measurement $|a\rangle=a_0|H\rangle+a_1e^{i\theta_a}|V\rangle$ and $|b\rangle=b_0|H\rangle+b_1e^{i\theta_b}|V\rangle$ should be determined by the relations
\begin{equation}
\begin{array}{l}
(n-|\alpha|)\theta_a+|\alpha|\theta_b=(2m_1+1)\pi,\\[5pt]
(n-|\beta|)\theta_a+|\beta|\theta_b+|\beta|\pi=(2m_2+1)\pi,\\[5pt]
b_0^{|\alpha|}a_0^{n-|\alpha|}=b_1^{|\alpha|}a_1^{n-|\alpha|},\\[5pt]
b_1^{|\beta|}a_0^{n-|\beta|}=b_0^{|\beta|}a_1^{n-|\beta|},
\end{array}
\end{equation}
where $a_i, b_i$ are all reals, and we take $m_1=m_2=0$ for simplicity. All the QWPs are set at $45^\circ$, angles of HWP using in experiment for scenario selected are listed in Table \ref{tab:table1}.


In experiment, we measured expectations of projection operators $b_\alpha a_{\bar{\alpha}}$,  $\bar{b}_\beta a_{\bar{\beta}}$, and $\prod_{k=1}^Na_k$ (which are equal to probabilities of successful projection). The experimental results are illustrates in Table \ref{tab:table2}, showing that the coincidence probabilities obey what quantum mechanics predicts. As we can see in the table, the measurements $\prod_{k=1}^Na_k$ obtain results 1 of explicitly non-zero probabilities. For $n=3$, $p_{\mathrm{std}}(a_1a_2a_3) = 0.132(16)$, $p_{\mathrm{opt}}(a_1a_2a_3)=0.259(22)$; for $n=4$, $p_{\mathrm{std}}(a_1a_2a_3a_4) = 0.093(12)$, $p_{\mathrm{opt}}(a_1a_2a_3)=0.127(12)$, where subscripts std and opt stand for standard and generalized optimal case, all the results are within the 1 standard deviation of theoretical prediction. Probabilities of projections $b_\alpha a_{\bar{\alpha}}$ and $\bar{b}_\beta a_{\bar{\beta}}$ is small but still deviate from zero due to unavoidable experimental errors caused by high-order photon emissions and imperfections of the photonic components. Therefore, only results listed in table \ref{tab:table2} are not sufficient to comfirm generalized Hardy's paradox, more powerful tools are needed to analysis whether the data we measured are within maximal tolerance.

For the paradox of $[n;|\alpha|,|\beta|]$-scenario, corresponding generalized Hardy's inequality is introduced as
\begin{equation}
\begin{array}{r@{}l}
\mathcal{I}[n;|\alpha|,&|\beta|;x,y]=F(n;|\alpha|,|\beta|;x,y)p(a_{I_n})\\[5pt]
&-x\sum_\alpha p(b_\alpha a_{\bar{\alpha}}) -y\sum_\beta p(\bar{b}_\beta a_{\bar{\beta}})\leq 0,
\end{array}
\label{eqn:hardyineq}
\end{equation}
with  $x>0, y>0$, here we take $x=y=1$; $F$ can be calculated as:
$$
F(n;|\alpha|,|\beta|;x,y)=\min_{0\leq m\leq n}\Big[x\binom{m}{|\alpha|}+y\binom{n-m}{|\beta|}\Big].
$$
Violation of the inequality (\ref{eqn:hardyineq}) implies our test of generalized hardy's paradox is effective.

We calculate $\mathcal{I}[n;|\alpha|,|\beta|;1,1]$ corresponding to the conditions we selected.
For $n=3$, $I(3;3,1;1,1)=0.055(22)$, $I(3;2,2;1,1)=0.150(27)$; for $n=4$, $I(4;4,1;1,1)=0.074(13)$, $I(4;2,2;1,1)=0.159(25)$. For the same GHZ state, $\mathcal{I}$ of optimal generalized case is positive by more standard deviations than previous standard one, at the same time optimal generalized case obtains higher probability to witness paradox, which is consistent with theoritical calculation. 

\emph{Conclusion.---} In summary, we give the first experimental test of the Hardy-typed quantum nonlocality in the multipartite scenarios. We utilize 3- and 4-photon GHZ states based on high-brightness beamlike c-BBO to experimentally perform the test. Our results confirm that the conflicts between quantum mechanics and LHV theories, revealed by a generalized Hardy's paradox, are stronger than those in the standard extensions of Hardy's paradox. The 3- and the 4-qubit generalized paradoxes, in particular, are stronger than the original 2-qubit paradox. The overall efficiency of our detection device is limited,
therefore, we need to assume that the detection efficiency is independent of measurement settings, known as fair sampling assumption~\cite{fairsampling}. To achieve higher detection efficiency, one can employ superconducting-nanowire single-photon detectors (SNSPD)~\cite{loopholefree1,loopholefree2}. As a further justification for the Hardy-typed quantum nonlocality, Bell's inequalities derived from the paradoxes are shown to be violated in quantum theory.

\begin{acknowledgments}
This work was supported by the National Natural Science Foundation of China (Grant No.\ 11475089), the Chinese Academy of Sciences, and the National Fundamental Research Program. 
\end{acknowledgments}

\end{document}